# Reversible local strain engineering of WS$_2$ using a micro-mechanical spring

*Eric Herrmann, Zhixiang Huang, Sai Rahul Sitaram, Ke Ma, S M Jahadun Nobi, and Xi Wang\**

Department of Materials Science and Engineering, College of Engineering, University of Delaware, Newark, DE 19716, USA

ABSTRACT: Local strain engineering is a promising technique to tune the properties of two-dimensional materials at the nanoscale. However, many existing methods are static and limit the systematic exploration of strain-dependent material behavior. Here, we demonstrate dynamic and reversible control of local strain distributions in suspended trilayer tungsten disulfide (WS$_2$) via nanoindentation using a micro-mechanical spring patterned with nanoscale probes. Micro-photoluminescence measurements reveal that indentation using a ring-shaped probe induces a nearly uniform biaxial strain distribution accompanied by a reversible redshift of the neutral exciton peak, consistent with simulated strain magnitudes. We further show that the in-plane strain distribution is spatially programmable by engineering the probe geometry and present designs for inducing point-like, uniaxial, biaxial, and triaxial strain distributions. The presented platform enables substrate-free, repeatable local strain engineering in suspended 2D materials and provides a versatile tool for streamlining the investigation of strain-dependent phenomena.

Strain engineering has emerged as a powerful technique for tailoring the electronic and optical properties of two-dimensional (2D) materials. While global strain engineering methods such as substrate bending have been widely adopted to dynamically tune the bandstructure of 2D materials[1–4], local strain engineering is quickly gaining significant attention due to its applications in nanophotonics and quantum information science[5–7]. In addition to modifying the local electronic and optical properties, local strain gradients can give rise to novel phenomena, such as pseudomagnetic fields in graphene[8–11] and exciton funneling in transition metal dichalcogenides (TMDCs)[12–17]. Several methods have recently been developed to control local strain in 2D materials systems, including the transfer of 2D materials onto nanostructured substrates[18–20], nanoimprint lithography[21,22], and strain transfer using stressor layers[23–26]. While these approaches are effective in the local patterning of strain, they are inherently static and currently limit the ability to systematically explore strain-dependent phenomena. Dynamic and reversible approaches to local strain engineering are therefore essential for streamlined and controlled investigation of local strain effects in 2D materials systems.

Among the most promising techniques for inducing dynamic local strain are those that employ an atomic force microscopy (AFM) tip for nanoscale indentation of suspended 2D materials. This method is widely used to extract the mechanical properties of 2D materials by relating the applied force to the resulting deformation of suspended flakes[27–29]. More recent studies have adapted this technique to probe strain-induced changes in the excitonic behavior of tungsten diselenide ($WSe_2$), both in suspended configurations[16] and in wrinkles formed during transfer[30]. However, the material response to strain is dependent on a variety of factors such as magnitude, distribution, and orientation relative to the crystallographic directions. For example, the spectral signature of certain types of quantum emitters in hexagonal boron nitride (hBN) redshifts under uniaxial strain along the armchair direction and blueshifts under uniaxial strain along the zigzag direction[31], three-fold symmetric strain distributions aligned to the crystallographic directions of graphene give rise to spatially uniform pseudomagnetic fields[32], and second harmonic generation enhancement in $NbOI_2$ is strongly correlated to the amount of strain along the polar axis[33]. The radially-symmetric strain distribution imposed by a single AFM tip is thus insufficient to study strain distribution-dependent effects on material properties. A strain engineering platform that permits programmable, dynamic control of the local strain orientation and magnitude is therefore essential to map the full strain-response landscape of 2D materials.

We introduce a novel platform for dynamically and reversibly engineering local strain distributions in suspended 2D materials via nano-indentation. Central to this approach is the design and fabrication of a silicon-on-insulator (SOI) based micro-spring (MS) with patterned nanoscale probes at its apex. By controlling the probe geometry, we introduce an additional degree of freedom through which to tune the local strain distribution. We demonstrate this concept using a ring-shaped probe to induce strain in suspended trilayer tungsten disulfide ($WS_2$) and measure the spectral shift of the neutral exciton peak across the strained region using spatially resolved micro-photoluminescence (PL) measurements. Finite element simulations suggest that the ring probe induces nearly uniform biaxial strain in the region above the ring probe, consistent with the observed spatial distribution of the neutral exciton PL peak. We further show that various strain distributions can be designed within the finite element framework and provide examples of probes that can be used to induce point-like, uniaxial, and triaxial strain distributions. The capability to reversibly tune the local strain environment in suspended 2D materials introduces a methodology for streamlining the study of strain-induced phenomena while avoiding complications that may arise due to substrate interactions.

A scanning electron microscope (SEM) and optical microscope images of a fabricated MS system are shown in Figures 1a and b, respectively. Multiple criteria are used to develop the MS design and are discussed in Supporting Information Section I. The MSs are fabricated using commercially available SOI wafers (WaferPro) consisting of a 220 nm thick Si device layer atop a 3 μm thick buried thermal oxide (BOX) layer. The MS is composed of multiple concentric Si arms connected to their nearest neighbors by Si bridges for stabilization and the outermost arm is connected to a 50 μm wide anchor region to fix the bottom of the MS to the substrate. Chromium (Cr) stressor layers are strategically patterned across the Si surface at the bridge locations and provide interfacial stress that induces out-of-plane deformation of the MS upon sacrificial etching of the underlying BOX layer, similar to the behavior of bimorph actuators used in micro-mirror arrays[34,35]. A 3 μm diameter SiO$_x$ ring probe with 100 nm width at the apex is then fabricated at the center of the MS using a process described previously[36]. After release of the MS via sacrificial etching, the ring probe is located at the point of maximum deflection. The fabrication of the MS system is discussed in detail in Section II of the Supporting Information.

The design and optimization of the MSs are guided by finite element based structural mechanics simulations in COMSOL Multiphysics. In the simulations, the deformation of the MS is approximated by applying a thermal expansion model to the Cr stressor layer and is calibrated experimentally using Cr/Si micro-cantilevers (see Supporting Information Section I). To calculate the spring constant of the MS, we apply a variable boundary load to the ring probe (inset of Figure 2a) and plot the downward deflection of the MS as a function of applied load in Figure 2b. At zero applied load, the point of maximum deflection is located at the center of the MS where the ring probe resides. The deformation of the MS scales linearly with applied load within the range of 0-46.5 nN as indicated by the red shaded region in Figure 2b. Within this range, the spring constant of the MS is $k = 0.0055$ N/m, comparable to AFM cantilever spring constants used to induce strain in suspended 2D materials in other works[16,37]. At a boundary load of 46.6 nN, the MS undergoes slight buckling which does not affect the location of maximum deflection. Between 46.6 and 111.2 nN, the deformation increases linearly with applied load, but with a slightly smaller spring constant of 0.0050 N/m. Above 111.2 nN, the MS undergoes a buckling mode near the apex such that the ring probe is no longer the location of maximum deflection. For visualization purposes, the cross-sectional deformation of the MS is shown in Figure 2c for various boundary loads corresponding to the highlighted data points in Figure 2b. These results indicate that the maximum applicable load that the MS can apply to suspended 2D materials is ~111 nN before the MS undergoes significant buckling, at which point the Si arms contact the flake. While this buckling behavior may be avoided by altering the in-plane geometry of the MS system (see Supporting Information Section II), we utilize the onset of the buckling mode to place an upper bound estimate on the maximum applicable load applied to suspended trilayer $WS_2$.

Suspended trilayer $WS_2$ is prepared by mechanical exfoliation from bulk crystal and stamped directly onto a silicon substrate containing a square aperture of width 10 μm. To promote adhesion to the silicon surface, a thin encapsulating layer of hexagonal boron nitride (hBN) is picked up prior to transfer over the aperture as shown in the schematic diagram in Supplementary Figure S6a. In Figures S6b and c, we show optical microscope images of the transferred stack as seen from above and below the aperture substrate, respectively. After transfer, the aperture is integrated into a home-built alignment system mounted onto the stage of a micro-PL microscope (Figure S6d). The aperture substrate is placed at the center of a 5-axis kinematic optic mount, which is then attached to the microscope stage using a custom adapter. As shown in the schematic illustration in Figure 3a, the MS substrate is mounted onto a height-controlled (z-axis) nanopositioner located on the stage and moves independently of the aperture substrate. This experimental setup enables micro-PL measurements of the suspended flake during indentation with the MS system.

The inset of Figure 3a shows the aligned MS-WS$_2$ system as seen through the suspended WS$_2$ immediately before contact. For the initial indentation, the MS probe is brought into contact with the suspended WS$_2$ and careful consideration is taken to avoid the significant buckling mode of the MS. High resolution (250 nm step size) cross-sectional PL spectra are taken at room temperature using 532 nm continuous wave excitation along the dashed line through the middle of the aperture shown in the inset of Figure 3a. The trilayer WS$_2$ PL spectra exhibit two primary active regions, between 600-700 nm and 700-850 nm attributed to the *K*-point and indirect transitions, respectively[38], and we utilize the spectral position of the indirect transition to confirm that the prepared WS$_2$ flake is trilayer[39,40]. In Figure 3b, we plot the cross-sectional spectra as heatmaps over the wavelength range of 600-700 nm before (left) and during contact (right) between the MS probe and WS$_2$, respectively. Before contact, we observe higher intensity PL near the edges of the aperture compared to the center of the suspended region, which may be attributed to collection of the reflected emission by the objective. During contact, we observe increased PL intensity near the contact region, accompanied by a spectral shift of the PL signal. To describe the observed PL behavior during contact, we perform fitting of each of the cross-sectional spectra using four Gaussian curves and show an example fitting of unstrained trilayer WS$_2$ in Figure S8. Here we focus on the Gaussian peak near 625 nm, which is attributed to the neutral exciton of trilayer WS$_2$ [39–41]. By comparing the spectral location of the neutral exciton peak before and during contact, we measure a maximum redshift of ~12 meV near the contact region between the ring probe and suspended trilayer WS$_2$.

To assess the repeatability of the indentation and spectral shift, a total of three indentation cycles were performed. After each measurement, the nanopositioner was lowered by 65 μm to ensure full restoration of the MS to the unloaded state. The ring probe was then returned to the original contact position and additional cross-sectional PL profiles were obtained. The results of the three indentations are summarized in Figure 3c, which plots the neutral exciton peak position along the dashed cross-sectional line indicated in the inset of Figure 3a. The results demonstrate repeatable indentation and spectral response of the neutral exciton peak, with a maximum redshift of approximately -12 ± 2 meV observed near the contact region, given by the blue error band in Figure 3c.

Dynamic strain engineering experiments were performed by acquiring point spectra directly above the contact region throughout the approach and retraction of the MS during an indentation cycle. Figure 3d presents the spectral shift of the neutral exciton as a function of MS position, where a position of zero corresponds to the onset of MS retraction. A clear redshift is observed as the MS engages the $WS_2$, followed by a return toward the unstrained exciton energy upon MS retraction. These results indicate that the MS platform enables the dynamic and reversible application of strain in suspended $WS_2$. We note that the presence of scatter and hysteresis in the acquired data may be attributed to mechanical drift and instability of the micro-spring during indentation. Similar observations were made in the dynamic indentation of suspended monolayer $WSe_2$ using a conventional AFM tip [16]. Improvements to the experimental setup, such as air-tight enclosure and inclusion of a closed-feedback loop, may improve stability during data acquisition.

Lower resolution (500 nm step size) spatial PL mapping is then performed across the 10 μm × 10 μm suspended region during indentation. The spatial distribution of the fitted neutral exciton peak, associated intensity distribution at the peak center, and peak width are shown in Figure 4a.

The maximum PL shift is located near the contact region, as revealed by the cross-sectional spectra previously. We also note red-shifting of the neutral exciton peak along paths connecting the ring probe to the corners of the aperture. While the PL intensity distribution varies across the suspended region of the flake, the intensity is highest near the corners of the aperture, which may be due to increased collection of reflected emission as mentioned previously.

To understand the observed PL shift distribution more clearly, we perform finite element based structural mechanics simulations using COMSOL Multiphysics to model the strain distribution that arises during indentation (see Section III of the Supporting Information). The trilayer $WS_2$ is treated as a three-dimensional domain and the force exerted by the ring probe is approximated as a boundary load of 100 nN, near the maximum applicable force prior to the significant buckling mode of the MS. The total applied force is distributed over a ring-shaped region and, because the ring probe itself is not explicitly modelled, friction is not included in the simulations. In Figure 4b, we plot the simulated $WS_2$ deformation (left) under the applied load and the resultant strain distribution $\varepsilon_{xx} + \varepsilon_{yy}$ (right). Interestingly, the deformation of the $WS_2$ is approximately constant within the perimeter of the ring probe as shown in the cross-sectional deformation plot of Figure 4c (left). Within this region, the strain components $\varepsilon_{xx}$ and $\varepsilon_{yy}$ are approximately equal, as illustrated in the right panel of Figure 4c. The near equivalence of $\varepsilon_{xx}$ and $\varepsilon_{yy}$ indicate that $WS_2$ is approximately symmetrically biaxially strained at a magnitude of $\varepsilon_{xx} + \varepsilon_{yy} = 0.10$ % across a large area, limited mainly by the shape of the aperture. A recent study[42] reported an average neutral exciton redshift rate of -127 ± 4 meV/% strain in monolayer $WS_2$ under biaxial loading. While direct comparison with the results for trilayer $WS_2$ is not possible without independent calibration, we note that a recent investigation determined the neutral exciton shift rates of monolayer, bilayer, and trilayer $WS_2$ are comparable under compressive hydrostatic pressure[43]. The neutral exciton redshift of -12 ± 2 meV measured here is thus consistent with a biaxial strain magnitude of 0.09 ± 0.02%. This estimate aligns well with the simulated strain values shown in Figure 4b.

We now investigate the feasibility of tailoring the strain distribution in suspended $WS_2$ by varying the probe geometry. Figure 5 summarizes simulation results of $WS_2$ indentation using a 10 nm diameter tip, a pair of parallel ridges, and three-fold symmetric tips of 100 nm apex diameter. In all simulations, a total force of 100 nN is applied. Figures 5a-d show the simulated flake deformation, strain distribution, principal strain directions, and cross-sectional strain profile for each probe geometries. As shown in the left panel of Figure 5a, the 10 nm diameter tip induces a deformation of the suspended flake that is much greater than that for the 3 μm diameter ring probe due to the smaller contact area. The highly localized contact generates a peak biaxial strain of ~ 5% directly above the tip, as shown in the left panel of Figure 5b. In the left panel of Figure 5c, the red and green arrows represent the first and second principal strain directions, respectively, and indicate that the biaxial strain at the point of contact becomes increasingly uniaxial away from the tip. This behavior is accompanied by a significant strain gradient in the vicinity of the tip, as shown in the cross-sectional strain profile in the left panel of Figure 5d.

Interestingly, we find that the parallel ridges and three-fold symmetric probes in Figure 5 give rise to strain distributions with symmetry similar to that of the probe geometry. While the strain magnitude is largest in the region of the $WS_2$ directly above the contact surfaces, there exists finite strain between the probes as a result of sliding of the $WS_2$ across the probe apex. In the case of the parallel ridges, the strain distribution transitions from asymmetric biaxial above the ridges to a uniaxial distribution at the center of the suspended region, as shown in the middle panels of Figures 5c and d. Similarly for the three-fold symmetric probes, the center of the suspended region exhibits nearly-symmetric triaxial strain limited mainly by the geometry of the aperture. These results indicate that the induced strain distribution may be tailored by careful design of the probe geometry.

While controlling the probe geometry in this manner is experimentally feasible, as exemplified by the SEM images in the insets of Figure 5a, exploration of dynamic and reversible strain control using the probes shown in Figure 5 is currently limited by experimental setup. The current configuration employs custom 3D-printed mounts designed for integration with conventional micro-PL microscopes. Although this setup dramatically reduces cost, it requires mounting the aperture substrate onto a 5-axis kinematic optic mount using a thin plastic plate with a hole drilled at the center (see Figure S6d). While this setup enables the simultaneous indentation and characterization of 2D materials using a 50× objective, the combined thicknesses of the plastic plate and aperture substrate do not support the use of a small working distance 100× objective. Further customization of the mount may enable the use of a 100× objective, increasing the PL collection efficiency and dramatically reducing the time required for high resolution PL mapping of the strained flake. Encasing the experimental setup in an enclosure may also increase stability and reduce environmental noise during dynamic indentation. These modifications may facilitate the higher resolution mapping required for quantitative comparison of experimentally measured PL and Raman scattering measurements to simulated strain distributions for the various probe geometries presented in Figure 4.

In summary, we demonstrate a micro-spring platform for dynamic, reversible, and spatially programmable local strain engineering in suspended two-dimensional materials. The observed strain-dependent photoluminescence behavior of trilayer $WS_2$ under indentation with a ring-shaped probe is consistent with strain distributions obtained from finite element-based structural mechanics simulations. We further show that the in-plane strain distribution can be tailored by engineering the probe geometry and present designs for inducing point-like, uniaxial, biaxial, and triaxial strain distributions. Importantly, the reversible and repeatable strain response enables successive measurements on the same flake. This reversibility, along with the design flexibility of the micro-spring system, may allow for systematic investigation of multiple strain configurations using interchangeable probe geometries. While this work focuses on suspended trilayer $WS_2$, the platform may be extended to other two-dimensional materials, enabling flake-specific studies of strain-dependent effects, including those related to symmetry and orientation with respect to the crystallographic directions.

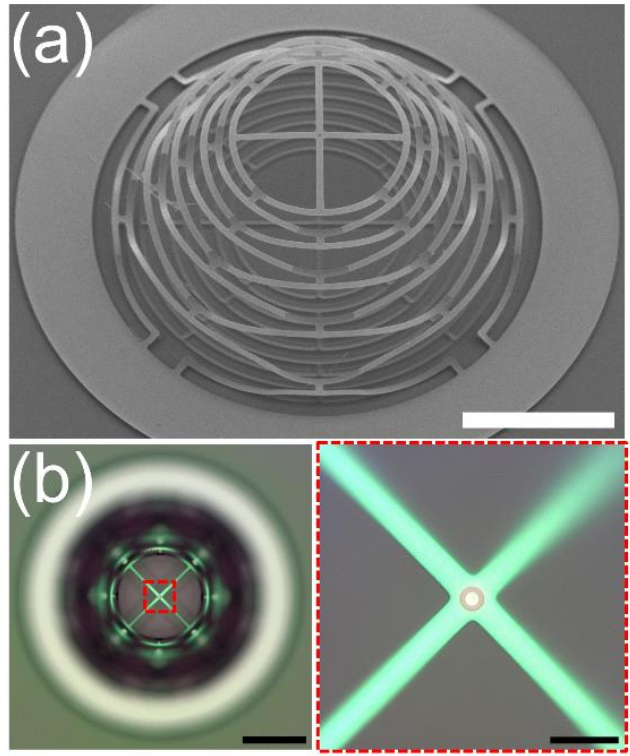

Figure 1. Fabricated micro-springs. (a) Tilted scanning electron microscope image of a fabricated micro-spring. The scale bar is 100 μm. (b) Optical microscope image of the micro-spring used in this work (left) and a zoomed-in image of the 3 μm diameter ring probe at the center of the spring (right). The scale bars in the left and right panels of (b) are 100 and 10 μm, respectively.

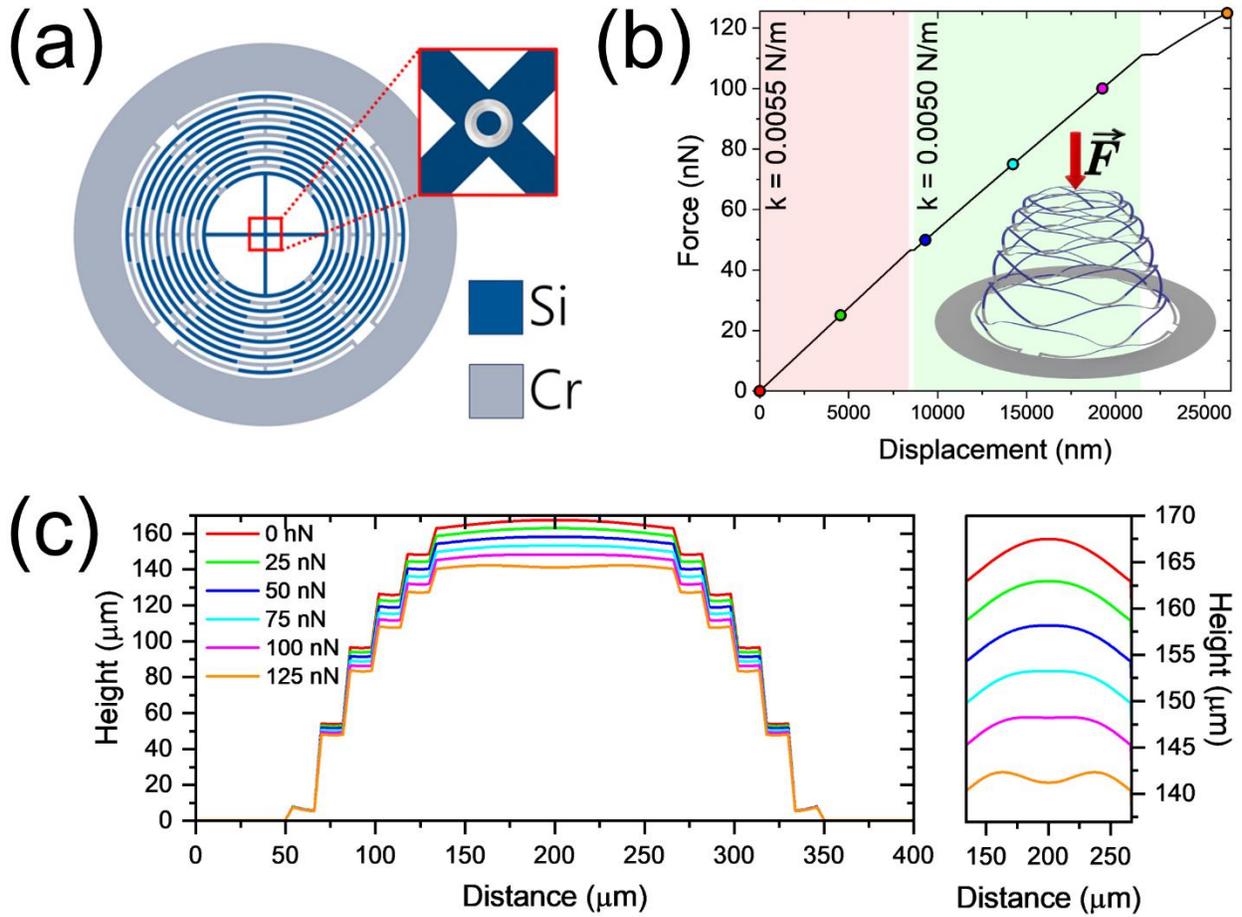

Figure 2. Micro-springs with patterned probes. (a) Schematic diagram of the micro-spring probe used in this work. Inset: close-up view of the 3 μm diameter ring probe. (b) Simulated spring constant of the micro-spring system. (c) Cross-sectional deformation profiles of the micro-spring under various applied loads corresponding to the highlighted data points in (b). The right panel shows an enhanced view of the deformation profile near the apex of the micro-spring. The left and right panels share the same legend.

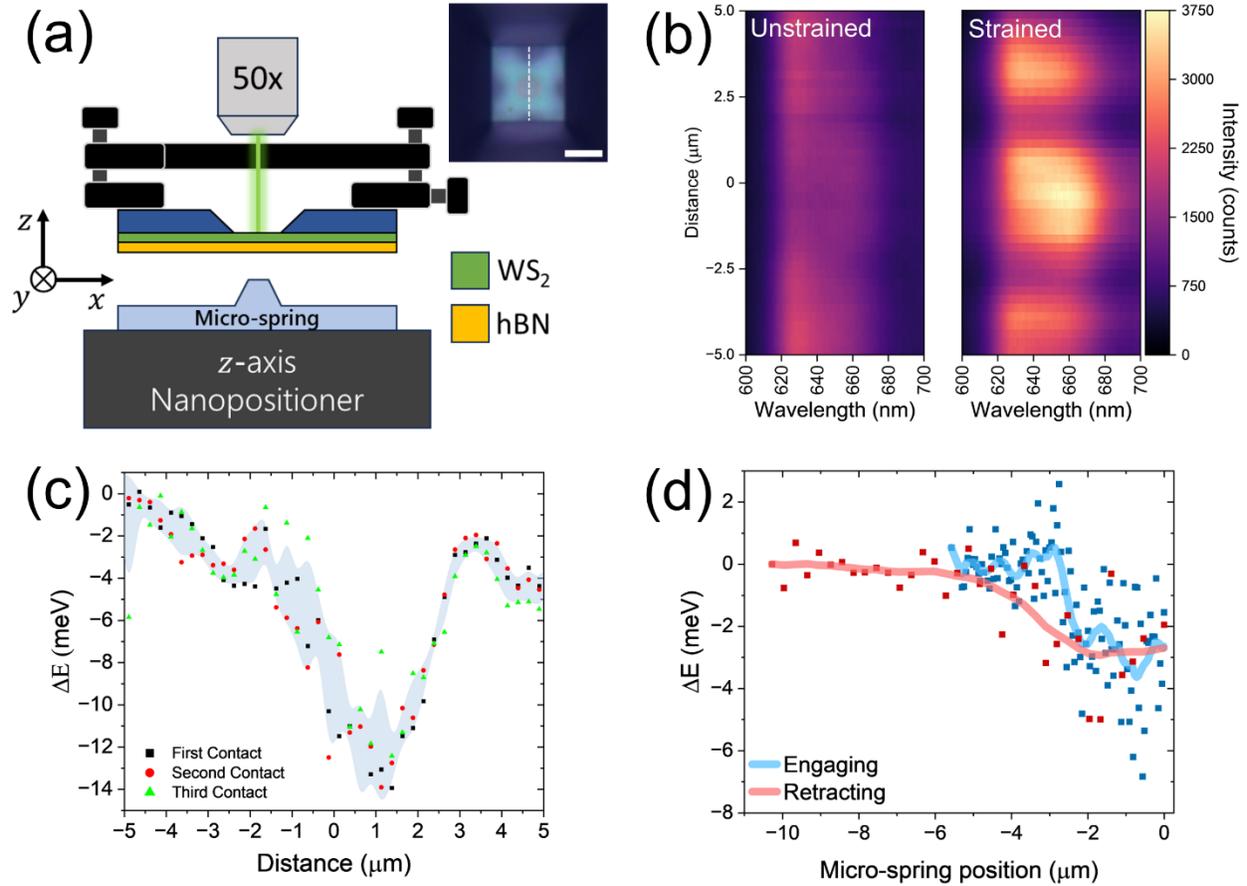

Figure 3. Reversible strain engineering. (a) Schematic diagram of the experimental setup. Inset: Optical microscope image of the suspended WS$_2$ as seen through the 50x objective immediately before contact between the micro-spring and suspended WS$_2$. The scale bar is 5 μm. (b) Cross-sectional photoluminescence spectra taken along the dashed line in the inset of (a) before (left) and during (right) indentation. (c) Spectral shift of the neutral exciton peak taken along the dashed line in the inset of (a) for three indentation measurements. The blue band shows the standard deviation of the measured peak shift. (d) Spectral shift of the neutral exciton peak during an indentation cycle using the micro-spring system. A position of zero corresponds to the onset of micro-spring retraction. The red and blue markers represent the neutral exciton spectral position during the engaging and retracting of the micro-spring, respectively. The solid curves are obtained by averaging adjacent data points (bin size of ten) and are included only as a visual aid.

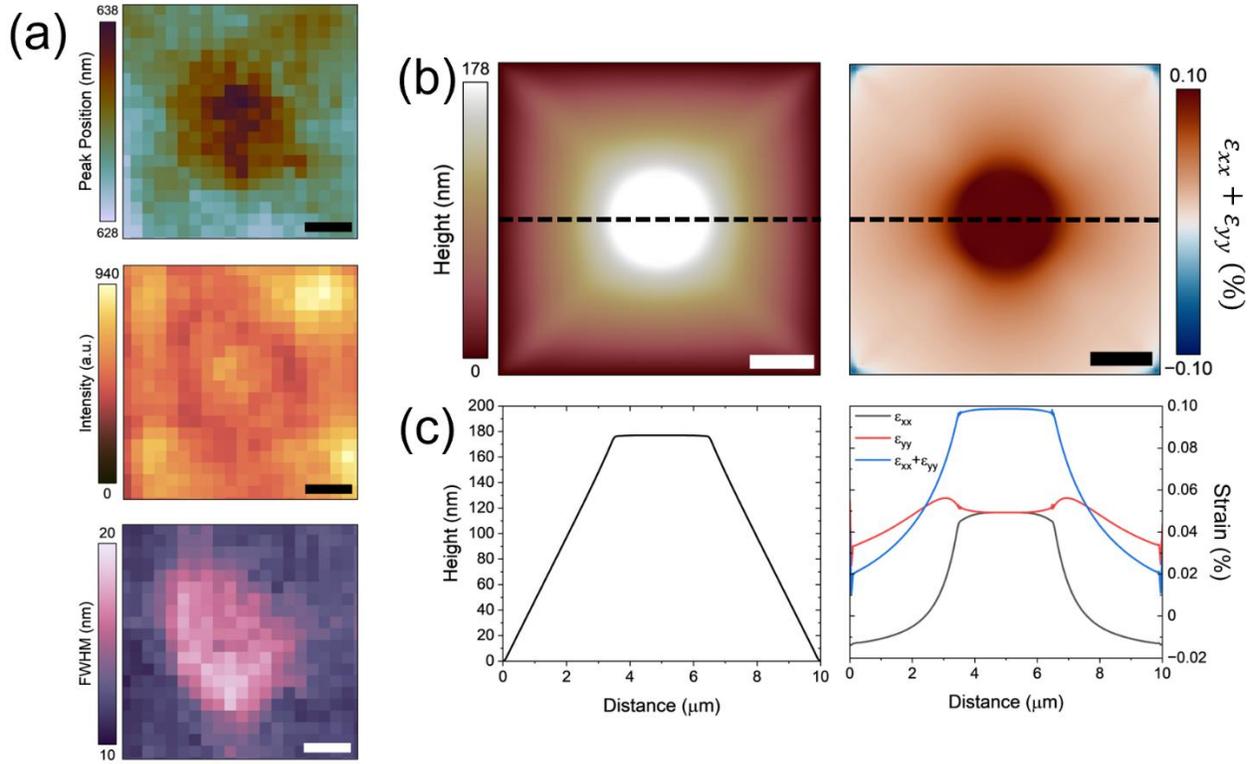

Figure 4. Indentation of suspended trilayer WS$_2$ using a 3 µm diameter ring probe. (a) Spatial distribution of the fitted neutral exciton peak emission (top), intensity distribution (middle), and peak width (bottom) across the extent of the suspended region. The scale bars are 2 µm. (b) Simulated deformation profile (left) and strain distribution (right) of the deformed flake. The scale bars are 2 µm. (c) Cross-sectional deformation (left) and strain (right) taken along the black dashed lines in (b).

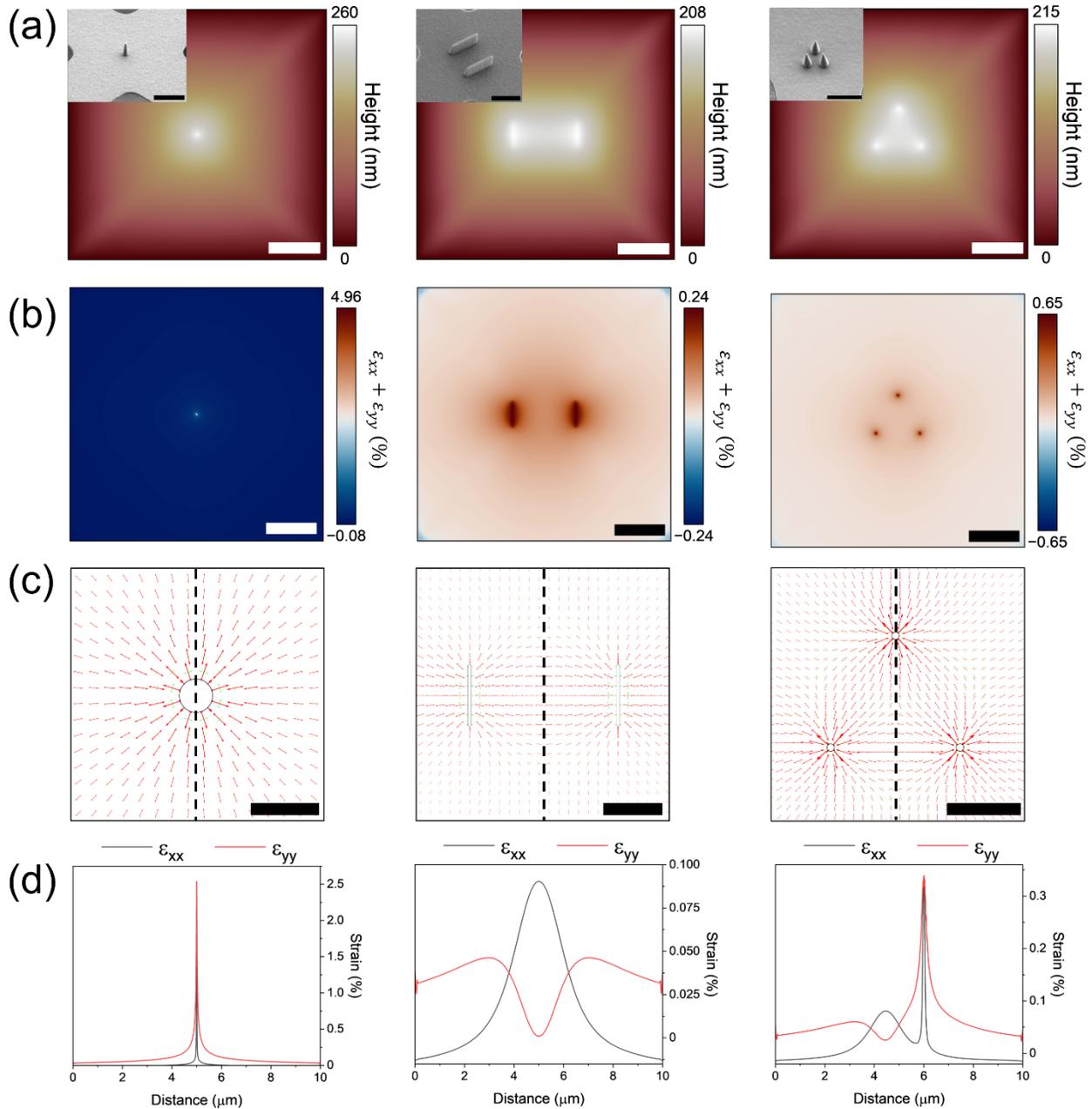

Figure 5. Tuning the strain distribution by probe control. (a) Simulated deformation, (b) strain distribution, (c) principal strain directions, and (d) cross-sectional strain along the vertical direction of trilayer WS$_2$ under indentation with various probes. The left, middle, and right columns correspond to indentation simulations using a 10 nm diameter tip, parallel ridges, and three-fold symmetric probes, respectively. All indentation simulations use a contact force of 100 nN. The cross-sections in (d) are taken along the black dashed lines in (c). The scale bars in the insets of (a) are 2 μm. The scale bars in (a) and (b) are 2 μm. The scale bars in the left, middle, and right panels of (c) are 20 nm, 1 μm, and 1 μm, respectively.

## ASSOCIATED CONTENT

**Supporting Information**.

The Supporting Information is available free of charge.

Micro-spring design and simulation, Micro-spring fabrication, Finite element simulations of strained $WS_2$, Experimental setup, Gaussian fitting of the photoluminescence spectra, Estimation of the force imparted by the micro-spring system.

## AUTHOR INFORMATION

**Corresponding Author**

**Xi Wang** − Department of Materials Science and Engineering, College of Engineering, University of Delaware, Newark, DE 19716, USA; Email: wangxi@udel.edu

**Author Contributions**

EH and XW conceived the idea and designed the experiment. EH performed mechanical simulations and fabrications with the support from ZH and SMJN. EH and SRS characterized fabricated micro-springs. EH, SRS, and KM prepared exfoliated 2D materials. EH performed optical characterizations. XW supervised the project. All authors participated in the discussion and contributed to refining the manuscript. All authors have given approval to the final version of the manuscript.

## ACKNOWLEDGMENT

This research was supported in part by the U.S. National Science Foundation under Grant DMR-2128534. E.H. acknowledges funding by the AG Microsystems MEMS Research Fund. S M J. N.

acknowledges the support by NASA Grant Number 80NSSC23M0076. Z. H. acknowledges the support by NSF through the University of Delaware Materials Research Science and Engineering Center, DMR-2011824. S. R. S. acknowledges the support by NSF under award ECCS-2102027.

**Supporting Information**

# Reversible local strain engineering of WS$_2$ using a micro-mechanical spring

*Eric Herrmann, Zhixiang Huang, Sai Rahul Sitaram, Ke Ma, S M Jahadun Nobi, and Xi Wang\**

Department of Materials Science and Engineering, College of Engineering, University of Delaware, Newark, DE 19716, USA

This PDF includes:

Section I: Spring design and calibration of the thermal expansion model
    Figures S1-S3
Section II: Device Fabrication
    Figure S4
Section III: Finite element simulations of strained WS2
    Figure S5
Section IV: Experimental Setup
    Figure S6
Additional Figures:
    Figure S7: Scanning electron microscope image of WS$_2$ suspended over a 10 μm square aperture.
    Figure S8: Fitting of the unstrained trilayer WS$_2$ photoluminescence spectrum
    Figure S9: Estimated uncertainty in the neutral exciton spectral fitting
    Figure S10: Estimation of the force imparted by the micro-spring in the experimental indentation measurement
    Figure S11: Optical microscope image of released micro-springs

**Section I: Spring design and calibration of the thermal expansion model**

    Multiple criteria were involved in the design process of the micro-spring (MS) and are summarized schematically in Figure S1. We choose to fabricate the MS using commercial silicon-on-insulator (SOI) wafers for ease of integration with existing fabrication techniques. The device requires probes with high aspect ratio to increase the separation between suspended flake and base of the spring. Taller probes minimize unwanted contact between the MS and aperture substrates that may arise due to fabrication irregularities or alignment error. Here, we find success using 1 μm tall probes. To strengthen the high aspect ratio probes under contact and avoid bending, we developed[1] and employ here a dry-etching procedure that enables control of the sidewall angle and supports sharp apex formation in $SiO_x$ nanostructures. Another crucial design consideration is the normal deflection of the MS. Unlike atomic force microscopy cantilevers that host only a single tip, here it is pertinent that the MS deflects normal to the plane of the substrate such that the force imparted onto the suspended $WS_2$ is uniform across the contact area. Finally, the MS itself must have a large deformation such that the flat regions of the substrate far away from the MS do not make undesired contact the aperture substrate. These design criteria.

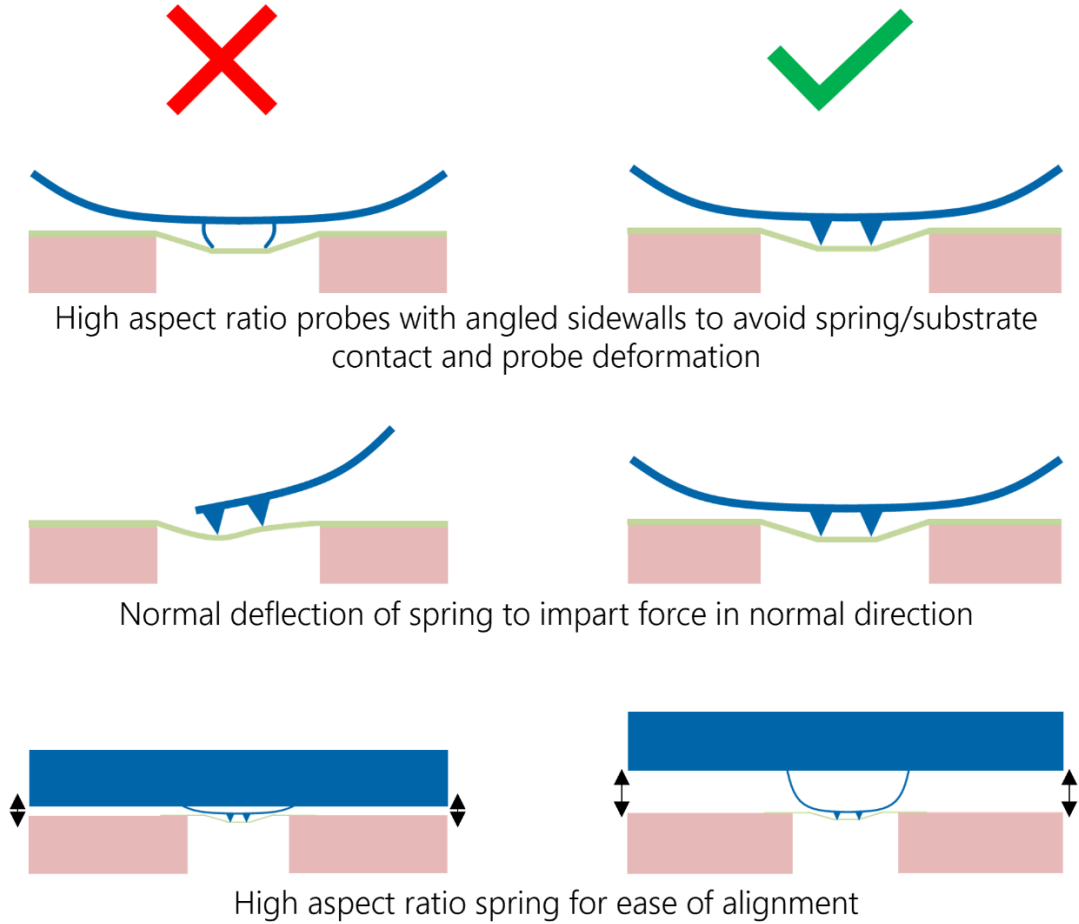

**Figure S1.** Micro-spring design criteria.

To realize these design criteria, we fabricate the MS using commercially available SOI wafers (WaferPro) with 220 nm Si device layer and 3 μm buried oxide (BOX) layer. The principle behind the deformation of the spring involves residual stress accumulation between the silicon (Si) surface of the MS and deposited chromium (Cr) stressor layers. This concept is widely employed as bimorph actuators used in the fabrication of micro-mirror arrays[2]. To describe the deformation of Si after patterning with Cr stressor layers, we perform finite element simulations using the Solid Mechanics module of COMSOL Multiphysics. While the origin of residual stresses in thin films is related to a variety of factors such as differences in thermal expansion coefficient, lattice mismatch, and impurities[3], we limit our model to thermal expansion alone for simplicity. In the simulations, the deformation is described using thermal expansion of the patterned Cr layers, where the Cr layers are cooled from an initial temperature $T_i$ to a final temperature of $T_f$ = 20°C. Here, $T_i$ represents the initial temperature of the Cr vapor as it begins to form a thin film across the MS surface during the electron beam evaporation process, a quantity that is difficult to measure experimentally. Calibration of $T_i$ is performed by first fabricating a series of Cr/Si micro-cantilevers and measuring the radii of curvature after release. A schematic illustration of the Cr/Si cantilevers is shown in Figure S2a. Figure S2b shows a scanning electron microscope (SEM) image of one series of fabricated cantilevers after release. Multiple series of cantilevers are fabricated across the sample and following release of the cantilevers, the sample is cleaved along a line through the pattern as shown by the red dashed line. Radii of curvature measurements are then taken using cross-sectional SEM as shown in Figure S2c. A series of five measurements yields an average radius of curvature of 25.2 ± 1.3 µm. The initial temperature $T_i$ is then varied in the simulations and the resultant deformation of the cantilever is shown in Figure S2d for various initial temperatures, ranging from 100-3000°C in steps of 100°C. The simulated radius of curvature is plotted as a function of initial temperature $T_i$ in Figure S2e and we find that a value of $T_i = 1400$ °C provides the best agreement with a radius of curvature of 25.3 µm.

We thus find that the deformation behavior of the micro-cantilevers is well-described by a thermal expansion model where $T_i = 1400$ °C. The thermal expansion model provides a method through which to estimate the residual stress accumulated within the Cr layer during the deposition process. Because deposition of the Cr stressor layers is carried out using identical conditions in the fabrication of the MS system, we use the calibrated thermal expansion model to simulate the deformation of the MS after release.

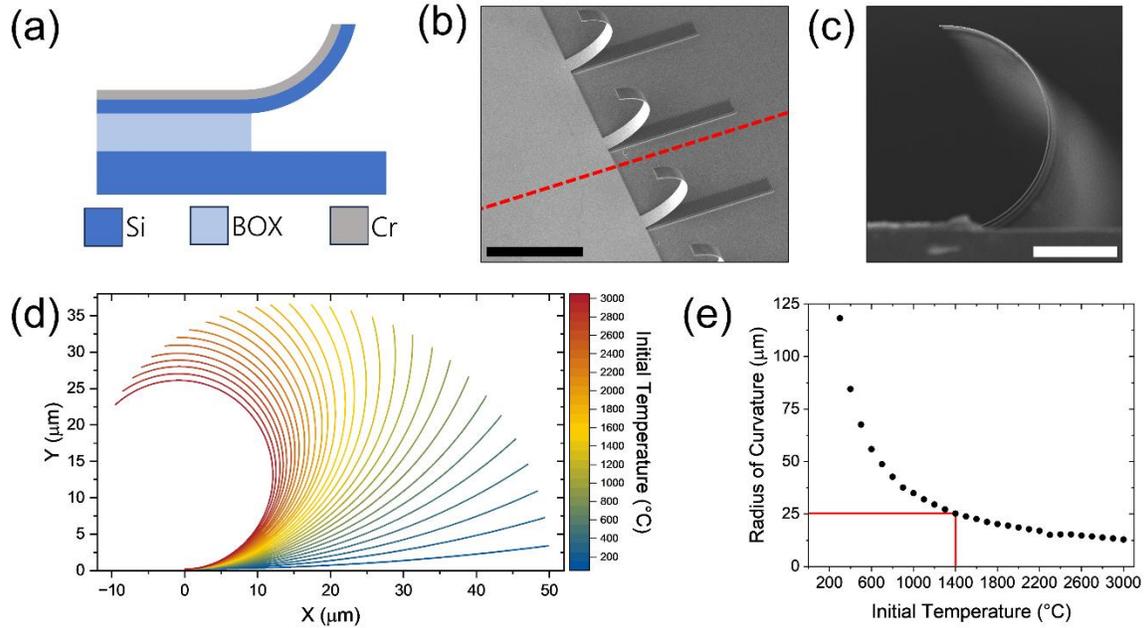

**Figure S2.** Calibration of the thermal expansion model. (a) Schematic illustration of the Cr/Si micro-cantilevers used in the calibration. (b) Scanning electron microscope image of fabricated micro-cantilevers and (c) their cross-sectional profile after cleaving the sample. The scale bars in (b) and (c) are 25 μm and 20 μm, respectively. (d) Simulated line profiles along the length of the micro-cantilevers as a function of initial temperature. (e) Extracted radius of curvature as a function of initial temperature.

To showcase the capability to tune the spring constant of the MS, we perform simulations for a slightly different version of the MS system with diameter 250 μm (MS$_{250}$), where the MS system discussed in the main text has a diameter of 300 μm (MS$_{300}$). In Figure S3a, we show a schematic illustration of MS$_{250}$ and in Figures S3b and c show the simulated force-displacement curves and cross-sectional deformation under various loads, respectively. Here, we find that MS$_{250}$ undergoes a minor buckling mode at 57.75 nN and the spring constant (0.012 N/m) is double that of the MS$_{300}$ presented in the main text.

Furthermore, MS$_{250}$ can withstand a force of 161 nN before the deformation becomes nonlinear with applied load, nearly a ~50% increase compared to MS$_{300}$. These results indicate that the spring constant and maximum exerted force may be tuned by changing the in-plane geometry of the MS. While this tunability is true for other spring systems such as atomic force microscopy cantilevers, the MS system is fabricated using a streamlined bottom-up approach and a variety of springs can be fabricated all within the same sample chip simply by changing the lithography pattern.

To validate the thermal expansion simulation using the MS system, we perform three-dimensional (3D) mapping of MS$_{250}$ using a spinning-disk confocal microscope (Andor Dragonfly) in reflectance mode. In Figure S3d, we plot the 3D deformation map and include projections along the two directions in the plane of the substrate surface. In Figure S3e, we compare the simulated and cross-sectional deformation profiles and find excellent agreement between simulation and experiment, indicating that the deformation behavior of the MS is well-described by thermal expansion of the Cr stressor layers.

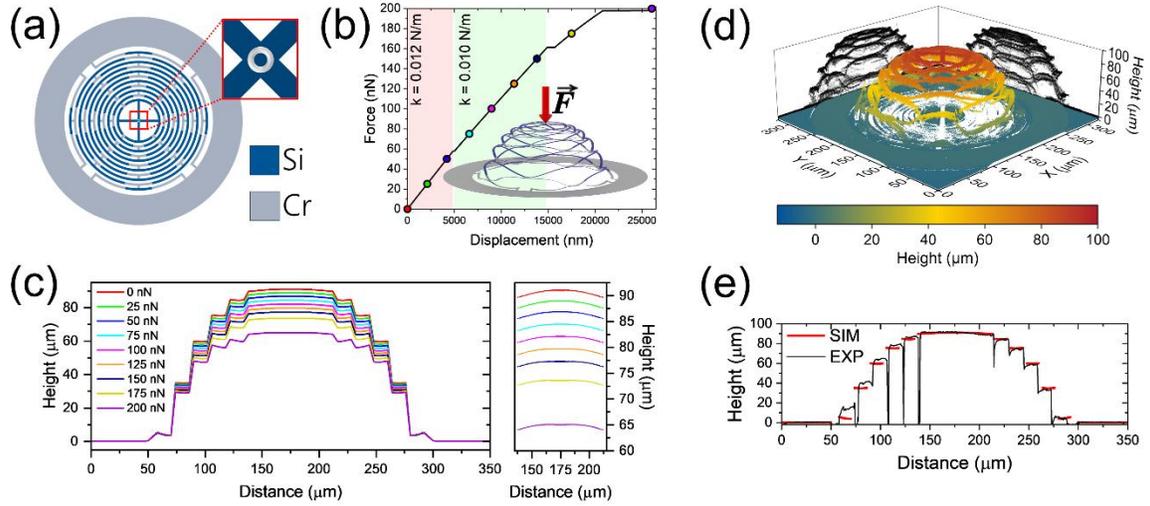

**Figure S3.** Micro-spring tunability and model verification. (a) Schematic illustration of a micro-spring with 250 μm diameter and 11 concentric arms. Inset: 3 μm diameter ring probe at the center of the micro-spring. (b) Simulated force-displacement curve for various forces applied to the apex of the ring probe. The two shaded regions indicate linear deformation. (c) Cross-sectional deformation profiles for various for various loads highlights in (b). (d) Three-dimensional mapping of the 11-arm spring probe. (e) Comparison of the simulated and experimentally measured cross-sectional deformation profiles.

**Section II: Device Fabrication**

The fabrication procedure is summarized in the schematic diagram of Figure S4. The fabrication of the MS begins with commercially-available silicon-on-insulator wafers (WaferPro) with a 3 µm thick buried oxide (BOX) layer and a 220 nm thick silicon (Si) device layer (Figure S4a). The base of the MS and the patterned 120 nm thick chromium (Cr) stressor layers are defined using conventional photolithography techniques. A 120 nm thick chromium (Cr) hard mask is then deposited at 0.5 Å/s using electron beam evaporation (PVD Products). After liftoff of the excess Cr, the spring base is vertically dry etched using a fluorine gas inductively coupled plasma system (PlasmaTherm) as shown in Figure S4b. The Si device layer is etched in an atmosphere of $SF_6$ and $C_4F_8$ with flow rates of 25 sccm and 20 sccm, respectively. The chamber pressure is held at 6 mTorr, the RF power is held at 50W, and the ICP power is held at 800W. The entirety of the Si device layer is etched through, plus an additional ~ 200 nm of the BOX layer. The process is designed to minimize the etch depth into the BOX layer so that conformal coverage of electron beam resist is attainable in subsequent steps.

The Cr hard mask is removed by immersion into a commercially-available Cr etchant solution held at 40°C for 30 minutes. Once the Cr hard mask is stripped, the sample is immersed in 6:1 buffered oxide etchant (BOE) for 8 minutes and 25 seconds. This step functions to begin an undercut of the BOX layer underneath the springs. This initial undercut of ~1.5 µm is necessary to ensure that the final spring release etch time is minimized while maintaining the integrity of the spring in subsequent steps. The patterned Cr strain patches are then defined using the same photolithography process that was used for defining the spring base (Figure S4c).

The probe fabrication begins with the deposition of a 1 μm thick $SiO_x$ layer across the entirety of the sample using plasma-enhanced chemical vapor deposition (PlasmaTherm) as shown in Figure S4d. This layer will be etched to define 1 μm tall $SiO_x$ probes. A bilayer resist stack consisting of LOR3A lift-off resist and AR-P 6200.09 positive-tone electron beam resist is then spun on top of the $SiO_x$ layer. The probe hard masks are then defined using electron beam lithography with a probe current of 1 nA and dose of 250 μC/cm². After developing the exposed resist, a 30 nm thick Cr hard mask is deposited at 0.2 Å/s using electron beam evaporation (Figure S4e). The Cr hard mask patches will define the in-plane geometry of the probes. The $SiO_x$ is then etched using fluorine gas inductively coupled plasma etching using a dual gas mixture of $CHF_3$ and $C_4F_8$ held at flow rates of 30 sccm and 40 sccm, respectively. The chamber pressure is held at 8 mTorr, the RF power at 30W, and the ICP power at 300W (Figure S4f). The etch rate is ~0.95 nm/s. This etch procedure is discussed in detail in a prior work[1].

After the probes are defined, a 5 μm thick photoresist layer (AZ P4330-RS) is spun onto the sample. The resist is then patterned into cylinders at the center of the MS to protect the probes during final release of the MS (Figure S4g). The springs are then immersed in BOE for an additional 10 minutes to release the springs from the substrate. After release, the BOE bath undergoes a dilution process in preparation for critical point drying. An intermediate step during dilution from BOE to isopropyl alcohol involves immersion of the sample into an N-methyl-2-pyrollidone bath for 30 minutes for removal of the photoresist protecting the probes. After removal from the critical point dryer (Figure S4h), the MS deflects upward out of the plane of the substrate as shown in Figure 1 of the main text.

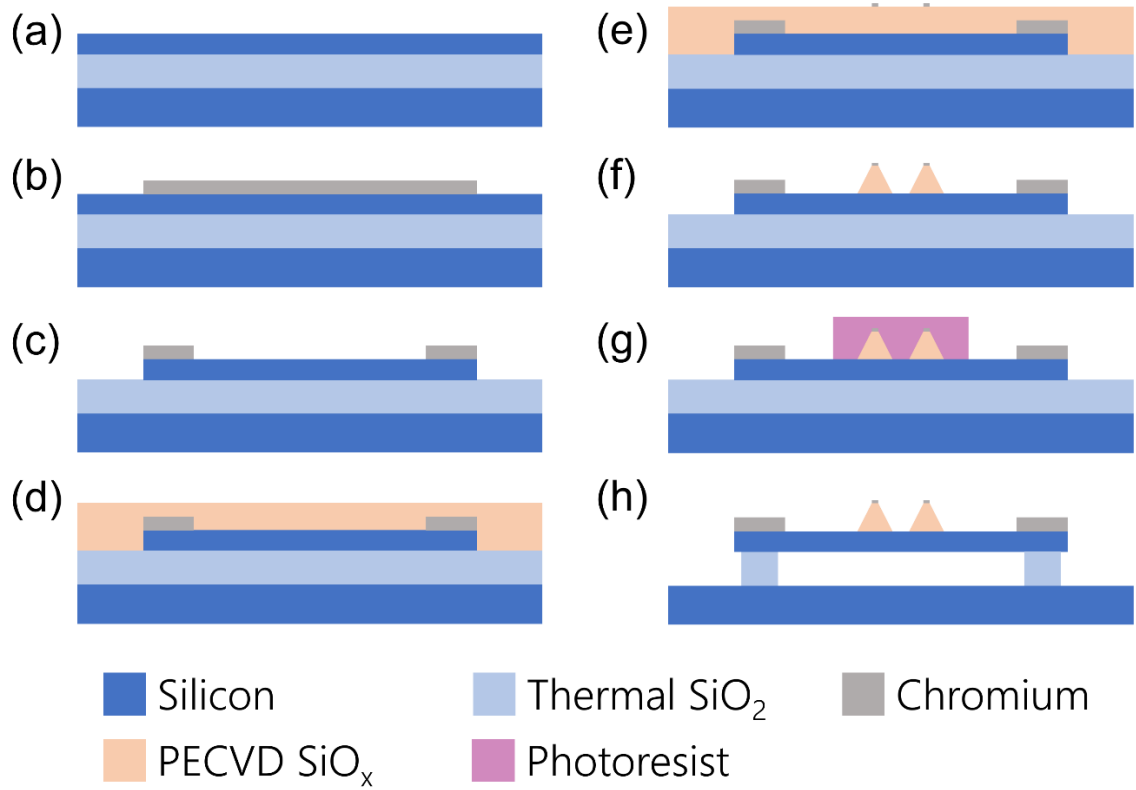

**Figure S4**. Schematic diagram of the micro-spring fabrication procedure.

**Section III: Finite element simulations of strained WS$_2$**

We performed finite element based structural mechanics simulations in COMSOL Multiphysics to approximate the induced strain in suspended trilayer WS$_2$ upon indentation with the various probes main text. In the simulations, the trilayer WS$_2$ is treated as a three-dimensional domain of 2 nm thickness with a Young's modulus and Poisson's ratio of 270 GPa and 0.22, respectively[4]. As a simplification, the encapsulating hBN layer is excluded from the simulations and indentation of the WS$_2$ is modelled by applying a boundary load of 100 nN over a region equivalent to the contact area on the underside of the WS domain. In this way, the probes are not explicitly defined in the simulations and frictionless contact is assumed. The strain distribution is then extracted from the simulations by taking the average through-thickness strain at each spatial location of the WS$_2$ domain.

In Figure S5, we show the effects of the aperture geometry on the simulated strain distributions. Figures S5a and b show the simulated strain distribution of WS$_2$ indentation over a 10 μm × 10 μm square aperture. These datasets are identical to those shown in the right panels of Figure 4b and c of the main text and are included here for comparison. Figures S5c and d show the simulated strain distribution for indentation of WS$_2$ over a circular aperture with diameter 10 μm. The simulation results show that the strain components $\varepsilon_{xx}$ and $\varepsilon_{yy}$ are equivalent in the region directly above the ring probe when the aperture is circular, but slightly asymmetric when the aperture has a square shape. These results show that the aperture geometry plays a small but non-negligible role in the simulated strain distribution.

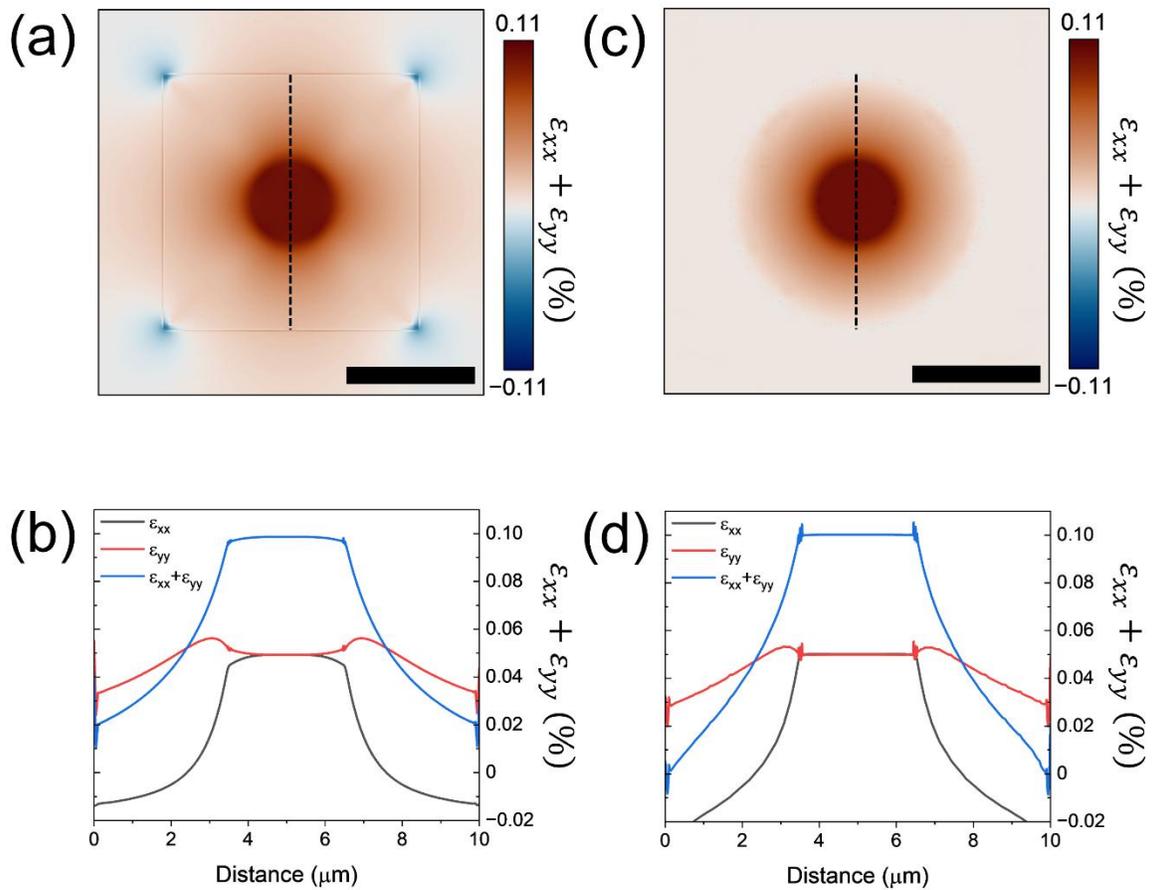

**Figure S5**. Influence of the aperture geometry on the resultant strain distribution. Simulated strain distributions for $WS_2$ over a (a) 10 µm × 10 µm square aperture and (b) a 10 µm diameter circular aperture. (b) and (d) show the strain components extracted along the black dashed lines in (a) and (c), respectively. The scale bars in (a) and (c) are 5 µm.

**Section IV: Experimental Setup**

In Figure S6a, we show a schematic illustration of the trilayer WS$_2$/hBN stack transferred over the 10 x 10 µm silicon aperture described in the main text. In Figures S6b and c, we show optical microscope images of the suspended flake as viewed from the indentation and optical characterization sides, respectively. The prepared sample is mounted at the center of a 5-axis kinematic optic mount and is attached to a micro-photoluminescence microscope stage using a custom 3D-printed adapter, as shown in Figure S6d. A separate 3D-printed adapter is used to mount the nanopositioner to the stage underneath the kinematic mount. This setup enables independent alignment of the micro-spring system to the suspended WS$_2$ using the kinematic mount tuning knobs.

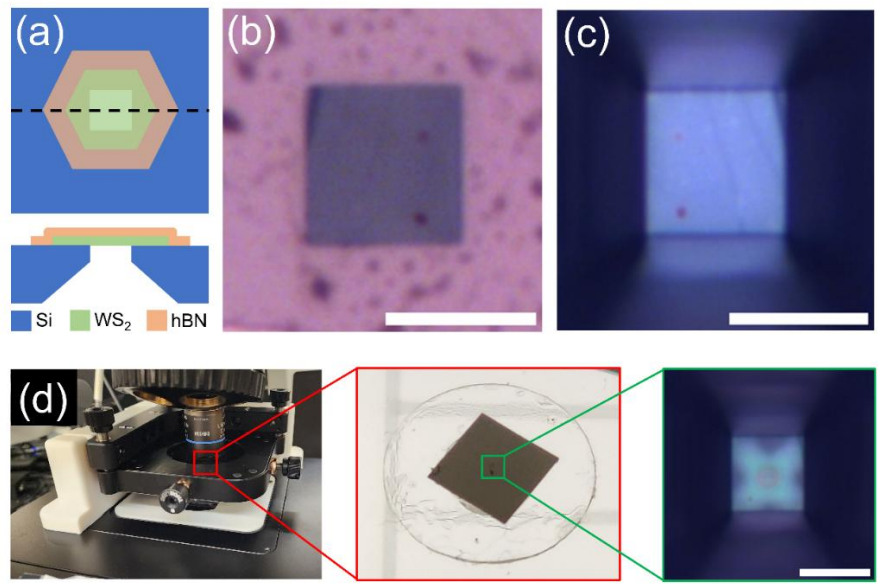

**Figure S6**. Experimental setup (a) Schematic diagram of hBN/WS$_2$ stack stamped onto a 10 µm wide silicon aperture as seen from above (top panel). A cross-sectional view taken along the dashed black line in the top panel is shown in the bottom panel. Optical microscope images of the transferred stack as seen from (b) above and (c) the underside of the aperture. (d) Custom 3D-printed mounts integrated onto the stage of a microscope. The insets show the location of the aperture (red) and an image of the micro-spring with ring probe aligned underneath the stack (green). All scale bars are 10 µm.

**Additional Figures**

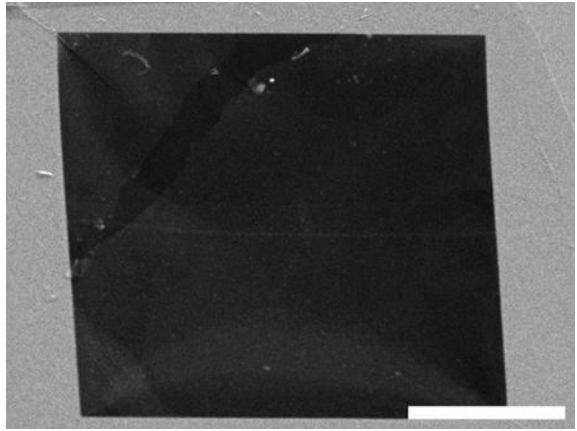

**Figure S7**. Scanning electron microscope image of a thick $WS_2$ flake suspended over a 10 μm x 10 μm aperture.

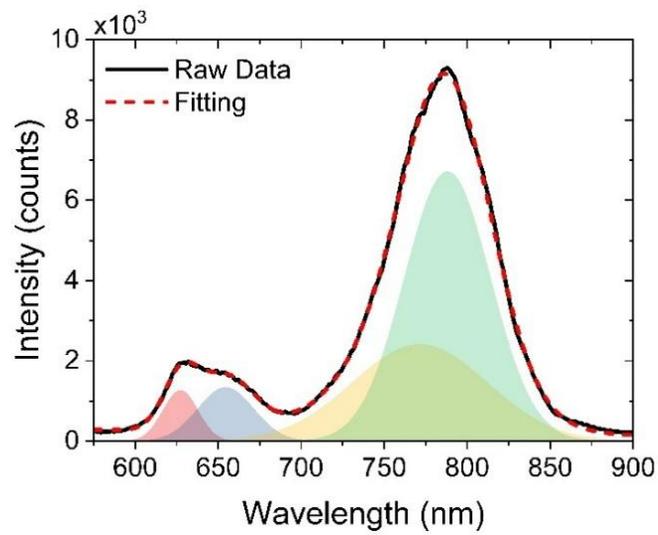

**Figure S8**. Fitting of the unstrained trilayer $WS_2$ spectrum using four Gaussian curves. The curve near 625 nm (red curve) is attributed to the neutral exciton peak[5–7].

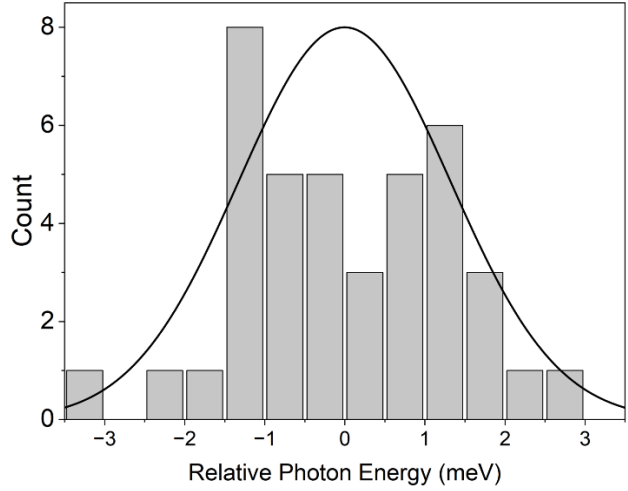

**Figure S9**. Estimation of the neutral exciton fitting stability. Distribution of unstrained trilayer $WS_2$ neutral exciton peak energy relative to the average for 40 independent spectra. The distribution exhibits a standard deviation of ±1.3 meV.

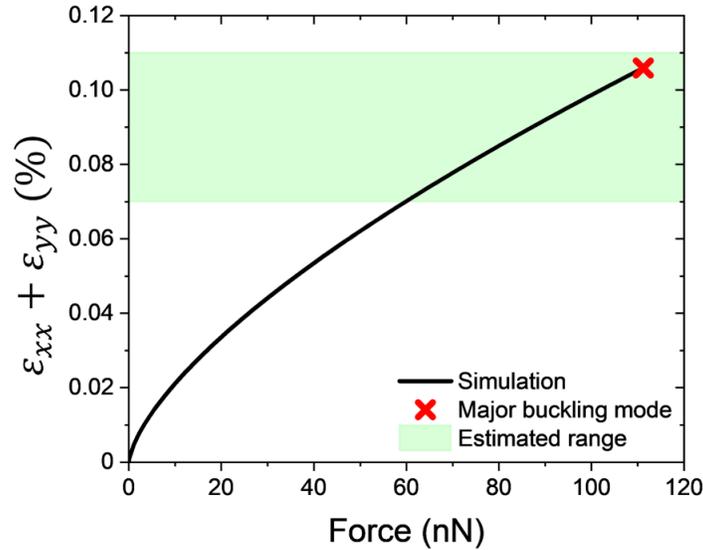

**Figure S10**. Estimation of the force imparted by the micro-spring in the experimental indentation measurement. The green shaded region corresponds to a biaxial strain magnitude of $\varepsilon_{xx} + \varepsilon_{yy} = 0.09 \pm 0.02\%$. This range is estimated using the neutral exciton redshift rate of -127 meV/% biaxial strain obtained from the literature[8] for monolayer $WS_2$ and the neutral exciton redshift rate measured in the main text for trilayer $WS_2$. We note here that the estimated strain using this method is an approximation. The estimated force imparted by the micro-spring in the main text is thus estimated to be between 60-111 nN, where the upper bound on the range is taken as the force at which the micro-spring undergoes the major buckling more, highlighted by the red cross symbol.

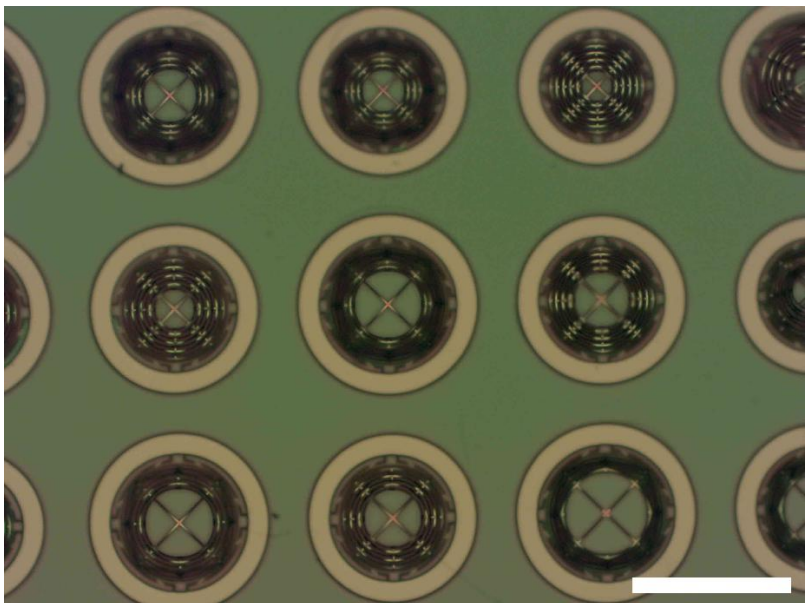

**Figure S11**. Optical microscope image of released micro-springs demonstrating the normal deflection and consistency of release. The scale bar is 300 μm.